
\magnification=\magstep1
\overfullrule=0pt\pageno=1
\hsize=15.4truecm
\line{\hfil OCIP/C-93-5}
\line{\hfil UQAM-PHE-93/07}
\line{\hfil hep-ph@xxx.lanl.gov/9306206}
\vskip.5cm
\centerline{\bf CONTRIBUTION OF SCALAR LOOPS}
\centerline{\bf TO THE THREE--PHOTON DECAY OF THE Z}
\vskip2cm
\centerline{HEINZ K\"ONIG\footnote*{email:Konig@osiris.phy.uqam.ca}}
\centerline{D\'epartement de Physique}
\centerline{l'Universit\'e du Qu\'ebec \`a Montr\'eal}
\centerline{C.P. 8888, Succ. Centre Ville, Montr\'eal}
\centerline{Qu\'ebec, Canada H3C 3P8}
\vskip2cm
\centerline{\bf ABSTRACT}\vskip.2cm\indent
I present the contribution of scalar particles to the decay of the
$Z^0$\ gauge boson into three photons, if their masses are larger
than the $Z^0$\ mass. As an underlying theory I take
the two Higgs doublet model and the minimal
supersymmetric standard model. I show that
the amplitude is identical to 0 up to the
second order in the expansion of the Higgs mass
squared. The maximal contribution
to the partial decay width of the $Z^0$\ to three photons via a
charged Higgs scalar loop is therefore about
$4\times10^{-16}\ (m_{Z^0}/m_{H^+})^8$\ GeV,
which is more than a factor of $5\times 10^4$\
 less than the W boson
contribution and so more than 6 orders of magnitude less than
the fermionic contribution. If we take into
account all families of the scalar partners of the leptons and quarks
this value is enhanced by a factor of about 50 and therefore still much
less as the W boson contribution and the fermionic one.
\vskip1cm
\centerline{ April 1993, revised version November 1993}
\vfill\break
{\bf I. INTRODUCTION}\hfill\break\vskip.2cm\noindent
There has been recent interest in the rare Z to three-photon decay
 [1-7 and references therein].
It is well known that due to Bose statistics and angular momentum
conservation the two-photon Z decay is forbidden (Yang-Mills theorem).
Whereas the three-photon Z decay is allowed, it however only occurs
in the standard model (SM) via one loop diagrams . In the SM there
are only two contributions; one with fermions in the loop and
the other one with W bosons. The first contribution was considered in
ref. [1-3]
and the second in ref. [4-6]. It was shown that the maximal fermionic
contribution to the partial decay width for $Z^0\rightarrow 3\gamma$\ decay
is about $7\times10^{-10}$\ GeV for a heavy top, and the W boson
contribution is only $2\times10^{-11}$\ GeV.\hfill\break\indent
Experimentally the $Z^0\rightarrow 3\gamma$\ decay  could be visible if its
branching ratio is larger than $10^{-5}$\ [6], which is certainly not the
case in the SM. Branching ratios of this order can be obtained by composite
models [6-7].\hfill\break\indent
Although the SM satisfies all presently known experimental
values there is strong belief in a model beyond the SM. One of the
most favoured models is the minimal supersymmetric standard
model (MSSM) [8-9]. The MSSM enhances the number of particles by at
least a factor of 2. One important new type of particles are
scalar particles such as the Higgs bosons and the
scalar partners of the leptons and quarks.
\hfill\break\indent
In this paper I present the contribution of the charged
Higgs particle to the three-photon Z decay and use as
an underlying model the two Higgs doublet model
(see e.g. [10] and references therein), but
because of the importance of the MSSM I also present
contribution of scalar quarks and leptons to this decay rate.
One purpose of this paper is to fill in the gap in calculating the
contributions of all three type (scalars, fermions and vector bosons) of
particles to this decay mode.\hfill\break\indent
In the next section I present the calculation. I discuss the
results in the Conclusions.
\hfill\break\vskip.2cm\noindent
 {\bf II. SCALAR PARTICLE CONTRIBUTION TO THE THREE-PHOTON Z DECAY\
}\vskip.2cm
There are four types of loop diagrams, which contribute to the
$Z\rightarrow 3\gamma$\ decay. They are shown in Fig.1.
 I start with the charged Higgs particle with its coupling
given in ref. [11] and then present the results of the contributions of scalar
quarks and leptons.\hfill\break\indent
In the calculation I include all terms and make no use
of gauge invariance,
that is I take $q^\mu, p_1^\alpha,
p_2^\beta, p_3^\gamma\not= 0$. The mass of
the photon I set to zero $p_i^2=0$. Here q is the
momenta of the $Z^0$\ boson, $\mu$\ its Lorentz index, $p_i$\ are
the momenta of the outgoing photons and $\alpha,\beta,\gamma$\
their Lorentz indices. Doing this we
are left with a large number of terms, which cannot be presented
in this short paper. Therefore I only comment on the calculation
procedure and give some results during the calculation, before
I present the final result.
\hfill\break\indent
After summation of all diagrams the logarithmic divergencies
(there are no higher ones) coming from dimensional integration
cancel out leaving us with a finite result. I show this explicitly below.
To present analytical
results, we have to do one simplification, otherwise the calculation
could be only done numerically from the beginning. We suppose that the
scalar particle mass is much larger than the Z boson mass. For the
charged Higgs and scalar quarks this is certainly a good approximation
in the MSSM. Only for the scalar leptons might it be false, although
the scalar neutrinos do not contribute to $Z\rightarrow 3 \gamma$\ decay
at all. Because we are only interested in the contribution scalar
particles might give to the three-photon Z decay I consider this
a reliable approximation.\hfill\break\indent
After a straightforward but tedious calculation it turns out
that the matrix element is identical to 0 in
the order $(m_{Z^0}/m_\Phi)^2$ when expanding in the scalar mass
and neglecting terms on the order $(m_{Z^0}/m_\Phi)^4$.
The reason is because of gauge
invariance. Writing down the amplitude in its most general form in
the order $(m_{Z^0}/m_\Phi)^2$ that is in the basis of $g_{\mu\alpha}
g_{\beta\gamma};g_{\mu\beta}g_{\alpha\gamma};g_{\mu\gamma}g_{\alpha\beta};
g_{\mu\alpha}\lbrace p_1,p_2,p_3\rbrace_\beta
\lbrace p_1,p_2,p_3\rbrace_\gamma;$\hfill\break
$g_{\mu\beta}\lbrace p_1,p_2,p_3\rbrace_\alpha
\lbrace p_1,p_2,p_3\rbrace_\gamma;
g_{\mu\gamma}\lbrace p_1,p_2,p_3\rbrace_\alpha
\lbrace p_1,p_2,p_3\rbrace_\beta;
g_{\alpha\beta}\lbrace p_1,p_2,p_3\rbrace_\mu\lbrace p_1,p_2,p_3
\rbrace_\gamma;$\hfill\break\noindent
$g_{\alpha\gamma}\lbrace p_1,p_2,p_3\rbrace_\mu
\lbrace p_1,p_2,p_3\rbrace_\beta;
g_{\beta\gamma}\lbrace p_1,p_2,p_3\rbrace_\mu
\lbrace p_1,p_2,p_3\rbrace_\alpha$\
(here the momenta in the parenthesis take all possible
combinations in the indices)
and using gauge invariance
$\lbrace p_1^\alpha,p_2^\beta,p_3^\gamma\rbrace\cdot M_{\mu\alpha\beta
\gamma}=0$\ shows that all coefficients in front of the upper terms
have to be 0. It is not possible to have gauge invariance and the
matrix element in the order $(m_{Z^0}/m_\Phi)^2$. Gauge invariance
is only possible if we also take into account all the
$\lbrace p_1,p_2,p_3\rbrace_\mu\lbrace p_1,p_2,p_3\rbrace_\alpha
\lbrace p_1,p_2,p_3\rbrace_\beta\lbrace p_1,p_2,p_3\rbrace_\gamma$\
terms. We are therefore forced to include the former neglected
$(m_{Z^0}/m_\Phi)^4$\ terms.
As a result the matrix element of the charged Higgs scalar
contribution to $Z\rightarrow 3 \gamma$\ decay can be written in the
following form
$$\eqalignno{iM=&\alpha^2\cot 2\Theta_W
{1\over 45}{1\over{m^4_{H^+}}}M^{\mu\alpha\beta\gamma}
\epsilon^q_\mu \epsilon^{*p_1}_\alpha\epsilon^{*p_2}_\beta
\epsilon^{*p_3}_\gamma&(1)\cr
M^{\mu\alpha\beta\gamma}=&A_1g^{\mu\alpha}g^{\beta\gamma}
+A_2g^{\mu\beta}g^{\alpha\gamma}+A_3g^{\mu\gamma}g^{\alpha\beta}
+B_1^{\beta\gamma} g^{\mu\alpha}
+B_2^{\alpha\gamma} g^{\mu\beta}+B_3^{\alpha\beta} g^{\mu\gamma}
\cr
&+C_1^{\mu\gamma}g^{\alpha\beta}+C_2^{\mu\beta}g^{\alpha\gamma}
+C_3^{\mu\alpha}g^{\beta\gamma}+D^{\mu\alpha\beta\gamma}\cr
A_1=&2\lbrack(p_1p_2)(p_1p_3)+3(p_1p_2)(p_2p_3)+3(p_1p_3)(p_2p_3)
\rbrack\cr
A_2=&A_1(p_1\leftrightarrow p_2)\cr
A_3=&A_1(p_1\leftrightarrow p_3)\cr
B_1^{\beta\gamma}=&-p_1p_2(2p_1+6p_2+p_3)^\gamma p_3^\beta
-p_1p_3(2p_1+p_2+6p_3)^\beta p_2^\gamma
\cr &+p_2p_3\lbrack
(2p_1+p_2)^\beta p_1^\gamma+p_1^\beta p_3^\gamma\rbrack\cr
B_2^{\alpha\gamma}=&B_1(p_1\leftrightarrow p_2,\beta\leftrightarrow
\alpha)\cr
B_3^{\alpha\beta}=&B_1(p_1\leftrightarrow p_3,\gamma\leftrightarrow
\alpha)\cr
C_1^{\mu\gamma}=&-p_1p_2(6p_1+6p_2+5p_3)^\gamma p_3^\mu
-p_1p_3\lbrack (-p_1+p_2+p_3)^\mu p_2^\gamma+p_2^\mu p_3^\gamma
\rbrack\cr &-p_2p_3\lbrack (p_1-p_2+p_3)^\mu p_1^\gamma+p_1^\mu
p_3^\gamma\rbrack\cr
C_2^{\mu\beta}=&C_1(p_2\leftrightarrow p_3,\gamma\leftrightarrow
\beta)\cr
C_3^{\mu\alpha}=&C_1(p_1\leftrightarrow p_3,\gamma\leftrightarrow
\alpha)\cr
D^{\mu\alpha\beta\gamma}=&p_1^\alpha(5p_1^\mu p_2^\gamma p_3^\beta
+p_1^\beta p_2^\gamma p_3^\mu+p_1^\gamma p_2^\mu p_3^\beta)
+p_2^\beta(p_1^\mu p_2^\gamma p_3^\alpha+p_1^\gamma p_2^\alpha p_3^\mu
+5p_1^\gamma p_2^\mu p_3^\alpha)\cr &+p_3^\gamma(p_1^\mu p_2^\alpha
p_3^\beta+p_1^\beta p_2^\mu p_3^\alpha +5 p_1^\beta p_2^\alpha
p_3^\mu)+q^\mu (p_1^\beta p_2^\gamma p_3^\alpha+p_1^\gamma p_2^\alpha
p_3^\beta)\cr &+6p_1^\mu p_2^\alpha p_2^\gamma p_3^\beta+
6p_1^\beta p_1^\gamma p_2^\mu p_3^\alpha+6(p_1^\mu p_2^\gamma+
p_1^\gamma p_2^\mu)p_3^\alpha p_3^\beta+6(p_1+p_2)^\gamma p_1^\beta
p_2^\alpha p_3^\mu\cr
q=&p_1+p_2+p_3\cr}$$
where the $\epsilon$\ are the polarization vectors of the Z boson and
photons.\hfill\break\indent
We have checked that the matrix element is gauge invariant in all
four momenta that is
$\lbrace q^\mu,p_1^\alpha,p_2^\beta,p_3^\gamma\rbrace
\cdot M_{\mu\alpha\beta
\gamma}=0$\ this is also true if we set $p_1^\alpha=p_2^\beta=
p_3^\gamma=0$.\hfill\break\indent
It is also worth making a few comments on the parameters $A_i$, because
here we can explicitly see how the divergencies cancel. Before the
expansion in terms of the scalar masses these parameters are given by
$$\eqalignno{A_i=&180\int\limits^1_0d\alpha_1\Bigl\lbrace \log f_i^1-
\int\limits_0^{1-\alpha_1}d\alpha_2\bigl\lbrace
2\log\limits_{j\not=k\not=i}f_{jk}^2+2\log f'^2_i-
\int\limits_0^{1-\alpha_1-\alpha_2}d\alpha_3\sum\limits^3_{i,j=1
\atop i\not=j}\log f_{ij}^3\bigr\rbrace\Bigr\rbrace\cr
f^1_i=&m^2_{H^+}-(p_i-q)^2\alpha_1(1-\alpha_1)\cr
f^2_{jk}=&m^2_{H^+}-2p_jp_k\alpha_1(1-\alpha_1-\alpha_2)\cr
f'^2_i=&m^2_{H^+}-q^2\alpha_1(1-\alpha_1)+2qp_i\alpha_1\alpha_2&(2)\cr
f^3_{ij}=&m^2_{H^+}-q^2\alpha_1(1-\alpha_1)-
2p_ip_j\alpha_2(1-\alpha_2-\alpha_3)
+2(p_i+p_j)q\alpha_1\alpha_2+2qp_i\alpha_1\alpha_3\cr
q^2=&m^2_{Z^0}\cr}$$
The $f$ functions would also be in the denominator of the terms
$B_i,C_i$\ and $D$ of $M^{\mu\alpha\beta\gamma}$\ in eq.1, but since we
neglect terms of the order higher than $(m_{Z^0}/m_{H^+})^6$\
we are only left with $m^4_{H^+}$\ in the denominator.
$f_i^1$\ comes from Fig.1 a), $f^2_{jk}$\ from Fig.1 b),
$f'^2_i$\ from Fig.1 c) and $f^3_{ij}$\ from Fig.1 d). After
integration over the Feynman parameters in eq.2 we explicitly
see that each constant term is cancelled ($1-{1\over 2}\cdot 2-
{1\over 2}\cdot 2+{1\over 6}\cdot 6\equiv 0$). This cancels
the logarithmic divergency in dimensional integration.
\hfill\break\indent
To calculate the partial Z decay width we have to square
the magnitude of eq.1.
I use the usual sum over polarizations for the photons and
for the Z boson and because of gauge invariance do
neglect the $q^\mu$ term there.
I also set $q_{\mu}\not=0$\ but $p_1^\alpha=p_2^\beta=p_3^\gamma=0$\
in the matrixelement. I do get the same result even without the last
condition.
As a result I obtain
$$\eqalignno{\vert M\vert^2=&{{8}\over{3\cdot 3!}}\Bigl({1\over{45}}\Bigr)^2
\alpha^4\cot^2 2\Theta_W\Bigl({{1}\over{m_{H^+}}}\Bigr)^8\Bigl\lbrace
68\Bigl\lbrack (p_1p_2)^2(p_1p_3)^2+(p_1p_2)^2(p_2p_3)^2+
\cr &(p_1p_3)^2(p_2p_3)^2\Bigr\rbrack+25m^2_{Z^0}(p_1p_2)(p_1p_3)
(p_2p_3)\Bigr\rbrace&(3)\cr}$$
We see that the final result is symmetric under the interchange of
different $p_i$'s.\hfill\break
The outcoming photons are massless and so we can even do the three body
phase space analytically. Because of the above mentioned symmetry it is
also not necessary to calculate every single term explicitely. In
the calculation of the three body phase space I make extensive use of
eq.B11 in [12], which can be
easily changed to the form needed here.
After this we are left with one final integration
over the energy of one outcoming photon, which has to be integrated from
$0$\ to $m_{Z^0}/2$.
\hfill\break\indent
As a final result I obtain for the partial three-photon Z decay width
$$\eqalignno{\Gamma_{H^+}(Z^0\rightarrow 3\gamma)=&
2.74\cdot 10^{-5}{{\alpha^4}\over{(4\pi)^3}}\cot^2 2\Theta_W\cdot
{{31}\over{80}} m_{Z^0}\Bigl({{m_{Z^0}}\over{m_{H^+}}}\Bigr)^8 &(4)\cr
\approx & 3.75\times 10^{-16}\Bigl({{m_{Z^0}}\over{m_{H^+}}}\Bigr)^8
\quad {\rm GeV}  \cr}$$
with $m_{Z^0}=91.17$, $\alpha=1/128$\ and $\sin^2\Theta_W=0.23$.
\hfill\break\vskip.2cm\noindent
{\bf III. CONCLUSIONS}\vskip.2cm
As a result we find that the charged Higgs scalar contribution
to the three-photon Z decay is always less than $3.75\times 10^{-16}$\ with
a further suppression factor of $(m_{Z^0}/m_{H^+})^8$, which cannot
be neglected as I did an expansion in terms of $(m_{Z^0}/m_{H^+})^2$ in
the calculation of the matrix element. The result is therefore
more than a factor
of $5\times 10^4$ less than the contribution of the W boson [4]
 and even more than
a factor
of $1.86\times 10^6$ less than the fermionic one [3].\hfill\break\indent
Finally I want to comment the contributions of the scalar partners
of leptons and quarks to the decay $Z^0\rightarrow 3\gamma$. The
calculation is exactly the same as for the charged Higgs scalar if
we make the same assumption that the masses of the scalar leptons
and quarks are much larger than the Z boson mass. In the final
result we just have to replace $\cot 2\Theta_W$\ by
$e^3_q(T_{3q}-e_q\sin^2\Theta_W)/\sin\Theta_W\cos\Theta_W$. Here
$e_q$\ is the charge and $T_{3q}$\ the isospin of the considered
particle. $\cot 2\Theta_W$\ suppresses the partial decay width
by a factor of $\cot^2 2\Theta_W\approx 0.41$. If we assume for
simplicity the masses of the scalar particles to be about the
same and sum over all the families (including the colour factor
for the quarks) we get an enhancement of a factor of
$\lbrack 3(1-44\sin^2\Theta_W/27)/\sin\Theta_W\cos\Theta_W\rbrack^2
\approx 19.86$\ which is a factor of 50 higher than for the charged
Higgs scalar giving us $1.82\times 10^{-14}$\ in eq.4.
As a final result we have that scalar particles lead in general
to a contribution to the three-photon Z decay width more than a
factor of 1000 smaller
as the W boson if the scalar masses are not too heavy.
\hfill\break\indent
In the MSSM we also know that the charginos (the mass eigenstates
of the fermionic partners of the charged Higgs scalar and W bosons)
contribute to the $Z^0\rightarrow 3 \gamma$\ decay similarly to
the fermions, which were considered previously [1-3].
In this paper we also did not consider interference terms as the
fermionic contribution is four orders of magnitude larger than
the scalar contribution. We only considered
the contribution of the scalar particles as one purpose of this paper
was to fill in
the gap in the calculation of the contributions of all kind of particles
to the three-photon Z partial decay width.
A general analysis within the MSSM will be presented elsewhere [13].
\hfill\break
\vfill\break\noindent
{\bf IV. ACKNOWLEDGEMENT}\vskip.2cm
I like to thank M. Baillargeon and C. Hamzaoui for very
helpful discussions. I am also grateful to M. Baillargeon for checking
eq.4 from the matrix element of eq.1.\hfill\break\indent
This work was partially funded by funds from the N.S.E.R.C. of
Canada and les Fonds F.C.A.R. du Qu\'ebec.
\vskip.2cm\noindent
{\bf REFERENCES}\vskip.2cm
\item{[\ 1]}M.L. Laurson, K.O. Mikaelin and A. Samuel, Phys.Rev.
{\bf D21}(1981)2795.
\item{[\ 2]}J.J. van der Bij and E.W.N. Glover, Nucl.Phys.{\bf B313}
(1989)237.
\item{[\ 3]}J.J. van der Bij and E.W.N. Glover, "Rare decays", in Z physics
at LEP 1, CERN 89-08 Vol.2 p.30.
\item{[\ 4]}M. Baillargeon and F. Boudjema, Phys. Lett.{\bf B272}(1991)158.
\item{[\ 5]}F. Dong, X. Jiang and X. Zhou, Phys.Rev.
{\bf D47}(1993)214, {\bf D46}(1992)5074.
\item{[\ 6]}D. Treille et al., Compositeness at LEP II,
CERN 87-08.
\item{[\ 7]}F. Boudjema and F.M. Renard, "Compositeness", in Z physics
at LEP 1, CERN 89-08 Vol.2 p.205.
\item{[\ 8]}H.P. Nilles, Phys.Rep.{\bf 110}(1984)1.
\item{[\ 9]}H.E. Haber and G.L. Kane, Phys.Rep.{\bf 117}(1985)75.
\item{[10]}J.F. Gunion et al., "The Higgs Hunter's Guide"
(Addison-Wesley, Redwood City, CA, 1990.
\item{[11]}J.F. Gunion and H.E. Haber, Nucl.Phys..{\bf B272}
(1986)1 (Fig.2 c) and d) and Fig.6 a) and b)).
\item{[12]} H. K\"onig, U. Ellwanger and M.G. Schmidt,
Z.Phys.-Particles and Fields {\bf C36}(1987)715.
\item{[13]}H. K\"onig (work in progress).
\hfill\break\vskip.2cm\noindent
{\bf FIGURE CAPTIONS}\vskip.2cm
\item{Fig.1}The diagrams with scalar particles
within the loop, which contribute to the three-photon Z decay.
There are 3 different ones in Fig.1 a), 6 in Fig.1 b)-d).
\vfill\break
\end